\documentclass[%
 reprint,
 amsmath,amssymb,
 aps,pra,
]{revtex4-1}

\usepackage{graphicx}
\usepackage{dcolumn}
\usepackage{bm}

\begin{document}

\preprint{APS/123-QED}

\title{Environment-induced synchronization of two quantum oscillators}

\author{Lo\"{i}c Henriet}
\affiliation{ICFO - Institut de Ciencies Fotoniques, Mediterranean Technology Park, 08860 Castelldefels (Barcelona), Spain}

\date{\today}

\begin{abstract}
Spontaneous synchronization between coupled periodic systems occur in a wealth of classical physical setups. Here, we show theoretically that the phase of two distinct quantum harmonic oscillators spontaneously  when they are strongly coupled to a common bosonic quantum dissipative environment at zero temperature, in the absence of any driving mechanism. To do so, we compute the dynamics of the oscillators with an exact master equation obtained from a path integral formalism. Above some value of the system-environment coupling strength, we observe numerically a strongly reduced damping and a frequency locking in the dynamics of the oscillators. Beyond the synchronization mechanism, we also describe the rich phenomenology of the dynamics in the model and notably identify the additional emergence of a regime with long-lived oscillations. 
\end{abstract}

\maketitle


\section{Introduction}

More than three centuries ago, Christiaan Huyguens observed that two nearby pendulum clocks attached to the same fixed basis would spontaneously synchronize (or anti-synchronize) due to their coupling to a common environment, and in the complete absence of any external
time-dependent driving force\,\cite{Huygens}. The emergence of such environment-induced synchronization effects in many different physical contexts encourages its investigation within quantum dynamical systems, for which there have already been numerous studies of synchronization in the presence of a driving force and nonlinearities\,\cite{Lee13,Lee14,Lorch17,Qiao18,Lorch17,Roulet18} or for quantum systems in contact with a common dissipative environment in the weak coupling regime\,\cite{Giorgi12,Manzano13,Bellomo17}. Recently, spontaneous synchronization was observed in the oscillatory dynamics of a cold atomic mixture of bosonic and fermionic species\,\cite{Delehaye2015}. The authors suggested that the appearance of this synchronization regime was due to the coupling of the relative motion of the two clouds with a common quantum dissipative environment, providing a quantum analog of Huyguens experiment.

In this article, we model this experimental result by considering a simple model of two quantum harmonic oscillators with different frequencies, that are coupled linearly to an infinite gaussian bosonic environment. By computing the exact dynamics of the system through a path integral approach, we show theoretically that the system enters a regime with phase-synchronization when increasing the coupling between the system and the environment $\alpha$, in agreement with experimental results. When $\alpha$ is increased further, we observe the emergence of a regime displaying long-lived phase-matched oscillations, that we relate to the non-equilibrium quantum phase transition described in Ref.\,\cite{Zhang2012}. We then present the dynamical phase diagram in the model.

\section{Model}

Our system is comprised of two harmonic oscillators, described by the annihilation (creation) operators $a_1$ ($a_1^{\dagger}$) and $a_2$ ($a_2^{\dagger}$), that have different frequencies $\omega_1\neq\omega_2$. We furthermore consider that the two modes are coupled to a quantum dissipative environment, leading to the following Hamiltonian
\begin{align}
\mathcal{H}=\sum_{j=1}^2 \omega_j a^{\dagger}_j a_j+\sum_{j,k} \lambda_{k,j} (a_j^{\dagger}b_k+b_k^{\dagger}a_j)  +\sum_k \omega_k b_k^{\dagger}b_k.\label{Ham}
\end{align}
Here, $b_k$ ($b_k^{\dagger}$) corresponds to the annihilation (creation) operator of the bosonic mode $k$ of the environment, with
frequency $\omega_k$ (we take $\hbar=1$). The interaction between the oscillator $j$ and
the mode $k$ is given by $\lambda_{k,j}$. Following Ref.  \cite{Delehaye2015}, we will suppose in the following that only the relative motion of the two clouds is coupled to the environment, i.e. $\lambda_{k,1}=-\lambda_{k,2}=\lambda_k$. The interaction between the oscillators and the environment is then fully characterized by the spectral function\,\cite{Leggett,weiss} $J(\omega)=\pi \sum_k \lambda_k^2 \delta(\omega-\omega)$, that we assume to be of ohmic form $J(\omega) =  \pi \alpha \omega e^{-\omega/\omega_c}$. Here, $\omega_c$ is a high frequency cutoff and the dimensionless parameter $\alpha$ quantifies the strength of the coupling between the system and the environment.
\begin{figure*}[t!]
\center
\includegraphics[width=1.\linewidth]{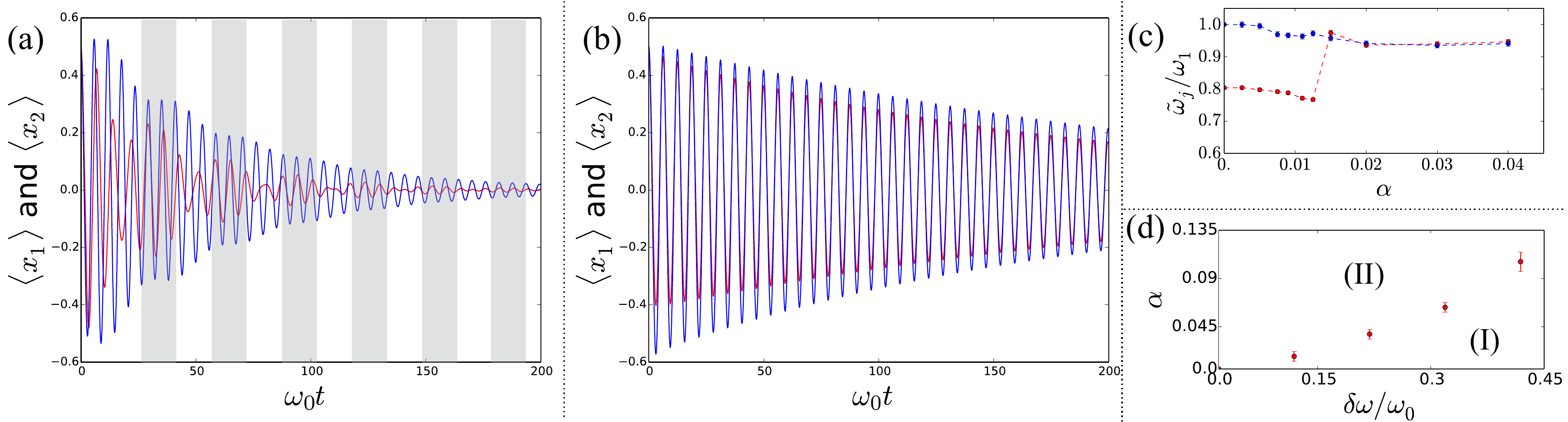}
\caption{(a) and (b) Dynamics of $\langle x_1 \rangle$ (in blue) and $\langle x_2 \rangle$ (in red) with respect to time for $\alpha=0.01$ (a) and $\alpha=0.16$ (b).  We have in both cases $\delta\omega/\omega_0=0.1$, $\omega_c/\omega_0=3$ and the two oscillators initially in a coherent state with initial values $\langle x_1 \rangle=\langle x_2 \rangle=0.5$. The shaded regions in (a) correspond to periods for which the oscillators are temporarily in phase. (c) Main observed frequencies in the dynamics $\tilde{\omega}_j/\omega_1$, obtained by locating the largest peak in the discrete Fourier transform of $\langle x_j \rangle$ (t), with respect to the coupling strength $\alpha$. Above a critical value, we observe a frequency locking of the oscillators. (d) Coupling strength above which the two frequencies of the oscillators are locked, as a function of the frequency mismatch $\delta \omega/\omega_0$.   }
\label{figure1}
\end{figure*}


\section{Exact master equation for the dynamics}

Computing the dynamics of Eq. (\ref{Ham}) beyond the weak coupling regime is in general not a straightforward task, as the coupling of the oscillators to the environment can induce strong Non-Markovian effects and entanglement in the dynamics \,\cite{Duarte06,Duarte09,McEndoo13,Charalambous18}. The quadratic form of the Hamiltonian makes it however possible to derive an exact master equation for the dynamics of the oscillators, by integrating out the quantum dissipative environment within a path integral formalism, as first presented in Ref.\,\cite{Hu92}. Let us present here the main steps of this derivation following this approach, that has subsequently been developed in Refs. \,\cite{Hu92,An07,An09,matisse_zhang_2,Zhang2012,Lin16}. It assumes factorized initial conditions between the system and the environment, i.e. $\rho (0) = \rho_{\mathcal{S}}(0)\otimes \rho_{\mathcal{B}}(0)$, where $\rho (0)$, $ \rho_{\mathcal{B}}(0)$ and $ \rho_{\mathcal{S}}(0)$ respectively denotes the total, bath-reduced and system-reduced density matrices at the initial time $t=0$. In this Letter, we will take a vacuum state for $\rho_{\mathcal{B}}(0)$, but the generalization to arbitrary initial temperature is straightforward.  Before presenting the main steps of the derivation, it is convenient to write the Hamiltonian in the center of mass and relative coordinates basis, $\psi_1=(a_1-a_2)/\sqrt{2}$ and $\psi_2=(a_1+a_2)/\sqrt{2}$. We obtain $\mathcal{H}=\mathcal{H}_0+\sqrt{2}\sum_k\lambda_k(\psi_1^{\dagger}b_k+b_k^{\dagger}\psi_1)+\sum_k \omega_k b_k^{\dagger}b_k$, with $\mathcal{H}_0=\sum_{j=1}^2 \omega_0 \psi^{\dagger}_j \psi_j+ \delta \omega(\psi_1^{\dagger}\psi_{2}+\psi_2^{\dagger}\psi_{1})$, where $\omega_0=(\omega_1+\omega_2)/2$ and $\delta \omega=(\omega_1-\omega_2)/2$.

The first step consists in writing the elements of the system density matrix at time $t>0$ within a coherent state path integral formalism. More specifically, $\langle \boldsymbol{\phi_f}  | \rho_{\mathcal{S}} (t)  | \boldsymbol{\phi_f'} \rangle$, where $ | \boldsymbol{\phi} \rangle=|\phi_{1},\phi_{2}  \rangle$ is a 2-mode coherent state in the center of mass and relative coordinates basis (i.e., with $\psi_j| \boldsymbol{\phi} \rangle=\phi_j| \boldsymbol{\phi} \rangle$), can be expressed as   
\begin{align}
\langle \boldsymbol{\phi_f}  | \rho_{\mathcal{S}} (t)  | \boldsymbol{\phi_f'} \rangle=\int d\mu(\boldsymbol{\phi_i})  d\mu(\boldsymbol{\phi_i'}) \langle \boldsymbol{\phi_i}  | \rho_{\mathcal{S}} (0)  | \boldsymbol{\phi_i'} \rangle\notag\\
\times J( \boldsymbol{\phi_f^*}, \boldsymbol{\phi_f'},t|\boldsymbol{\phi_i}, \boldsymbol{\phi'_i{}^*},0),
\label{time_evolution_general}
\end{align}
where $\mathcal{J}( \boldsymbol{\phi_f^*}, \boldsymbol{\phi_f'},t|\boldsymbol{\phi_i}, \boldsymbol{\phi'_i{}^*},0)$ is the total propagating function between $t=0$ and $t$ gathering forward and backward transition amplitudes \cite{FV,Leggett,weiss}. In Eq. (\ref{time_evolution_general}), $d\mu()$ denotes the standard integration measure in the coherent state path integral. We can express the propagating function $\mathcal{J}=\mathcal{J}( \boldsymbol{\phi_f^*}, \boldsymbol{\phi_f'},t|\boldsymbol{\phi_i}, \boldsymbol{\phi'_i{}^*},0)$ as a functional integral over the forward and backward 2-dimensional bosonic fields $\boldsymbol{\phi}$ and $\boldsymbol{\phi'}$ on the Keldysh contour,  
\begin{align}
 \mathcal{J}=\int \mathcal{D}[\boldsymbol{\phi},\boldsymbol{\phi^*},\boldsymbol{\phi'},\boldsymbol{\phi'^{*}}] e^{\mathcal{S}[\boldsymbol{\phi^*},\boldsymbol{\phi}]+\mathcal{S}^{\dagger}[\boldsymbol{\phi'^{*}},\boldsymbol{\phi'}]+\mathcal{M}}.
\label{propagating_function_general}
\end{align}
In Eq. (\ref{propagating_function_general}), $\mathcal{D}[\boldsymbol{\phi},\boldsymbol{\phi^*},\boldsymbol{\phi'},\boldsymbol{\phi'^{*}}]$ denotes the standard integration measure over the fields $\boldsymbol{\phi}$ and $\boldsymbol{\phi^*}$ with boundary conditions $\boldsymbol{\phi}(0)=\boldsymbol{\phi_i}$ and $\boldsymbol{\phi^*}(t)=\boldsymbol{\phi^*_f}$; and $\boldsymbol{\phi'^{*}}(0)=\boldsymbol{\phi'^{*}_{i}}$ and $\boldsymbol{\phi'}(t)=\boldsymbol{\phi'_{f}}$. The argument in the exponential contains the free forward and backward actions, with $\mathcal{S}[\boldsymbol{\phi^*},\boldsymbol{\phi}]=\boldsymbol{\phi^*}(t)\boldsymbol{\phi}(t)-\int_{0}^t d\tau \left[\dot{\boldsymbol{\phi}}(\tau)\boldsymbol{\phi^*}(\tau)+i \mathcal{H}_0[\boldsymbol{\phi^*},\boldsymbol{\phi}]\right]$.
It is important to note the presence of boundary terms in $\mathcal{S}$ that emerge naturally in the construction of the path integral. In the absence of the environment, the forward and backward fields would be uncoupled and the stationary trajectories could be determined independently. After the integration of the gaussian bosonic environment, they are here coupled through the additional  term $\mathcal{M}$, known as the Feynman-Vernon influence functional\,\cite{FV}, which reads
\begin{widetext}
\begin{align}
\mathcal{M}[\boldsymbol{\phi^*},\boldsymbol{\phi'^{*}},\boldsymbol{\phi},\boldsymbol{\phi}']=\int_0^t d\tau \int_0^\tau d\tau' \left[\phi_1'^*(\tau)-\phi_1^*(\tau) \right]\eta(\tau-\tau')\phi_1(\tau')+\left[\phi_1(\tau)-\phi_1'(\tau) \right]\eta^*(\tau-\tau')\phi_1'^*(\tau'),
 \label{FV_action}
\end{align}
\end{widetext}
with $\eta(\tau-\tau')= \sum_k \lambda_k^2 e^{-i\omega_k (\tau-\tau')}$. The expression of the spectral function given above leads here to $\eta(\tau-\tau')=2\alpha  \omega_c^2 /[1+i\omega_c (\tau-\tau')]^{2}$.

As the action is quadratic, the path integral can still be solved exactly using the stationary path method\,\cite{Klauder79}. One finds more precisely the stationary fields under the form $\phi_1(\tau)=u(\tau)\phi_{1i}+v(\tau)\phi_{2i}$, 
$\phi_2(\tau)=w(\tau)\phi_{1i}+x(\tau)\phi_{2i}$, $\phi'^*_1(\tau)=u^*(\tau)\phi'^*_{1i}+v^*(\tau)\phi'^*_{2i}$, 
$\phi'^*_2(\tau)=w^*(\tau)\phi'^*_{1i}+x^*(\tau)\phi'^*_{2i}$ where $u$, $v$, $w$ and $x$ are complex functions that verify  
\begin{align}
&\dot{u}(\tau)+i\omega_0 u(\tau) +i\delta \omega w(\tau)=-\int_0^{\tau}d\tau' \mu(\tau-\tau')u(\tau')\label{eq1}\\
&\dot{v}(\tau)+i\omega_0 v(\tau) +i\delta \omega x(\tau)=-\int_0^{\tau}d\tau' \mu(\tau-\tau')v(\tau')\label{eq2}\\
&\dot{w}(\tau)+i\omega_0 w(\tau) +i\delta \omega u(\tau)=0\label{eq3}\\
&\dot{x}(\tau)+i\omega_0 x(\tau) +i\delta \omega v(\tau)=0\label{eq4},
\end{align}
with initial conditions $u(0)=x(0)=1$ and $v(0)=w(0)=0$. 

The final step is to re-inject the stationary solutions into the path integral expression and differentiate the expression obtained from Eq. (\ref{propagating_function_general}) with respect to time. Using then the Bargmann representation of operators\,\cite{Vourdas94,Anastopoulos00}, we finally reach an exact master equation governing the dynamics of the density matrix of the two quantum oscillators, 

\begin{align}
\partial_t \rho_S=-i[\mathcal{H}_{\rm eff}(t)\rho_S-\rho_S \mathcal{H}_{\rm eff}^{\dagger}(t)]+\Gamma_1(t) \psi_1 \rho_S \psi_1^{\dagger}\notag\\
+\frac{\Gamma_2(t)}{2} (\psi_1 \rho_S \psi_2^{\dagger}+\psi_2 \rho_S \psi_1^{\dagger})\label{master}
\end{align}
where the time-dependent non-Hermitian Hamiltonian reads $\mathcal{H}_{\rm eff}(t)=i\frac{w\dot{v}-x\dot{u}}{wv-xu}\psi_1^{\dagger}\psi_1+i\frac{\dot{u}v-u\dot{v}}{wv-xu}\psi_2^{\dagger}\psi_1+\delta \omega\psi_1^{\dagger}\psi_2+\omega_0 \psi_2^{\dagger}\psi_2$, and $\Gamma_1/2=-\Im [i\frac{w\dot{v}-x\dot{u}}{wv-xu}]$, $\Gamma_2/2=-\Im [i\frac{\dot{u}v-u\dot{v}}{wv-xu}]$. Equation (\ref{master}) above is an exact, trace-preserving and time-local master equation describing the system degrees of freedom only, which can conveniently solved numerically. In practice, we will solve Eq. (\ref{master}) in a truncated space, by considering that at most $N_c$ excitations are present in the system, and subsequently assert the accuracy of the results obtained by checking that they are independent of $N_c$. It is important to note that Eq. (\ref{master}) encapsulates the non-Markovian effects on the dynamics induced by the environment through the time-dependency of $\Gamma_j$ and of the coefficients in $\mathcal{H}_{\rm eff}$.

\section{Environment-induced synchronization}

Armed with the exact master equation (\ref{master}), we now study the effect of the environment on the dynamics of the two oscillators. We first consider a frequency mismatch $\delta \omega/\omega_0=0.1$ between the two modes. At weak coupling $\alpha$, the environment induces an asynchronous relaxation of the two oscillator coherences towards the equilibrium value $\langle a_1 \rangle =\langle a_2 \rangle=0 $. This is illustrated in Fig.\,\ref{figure1}(a), where we show the dynamics of $\langle x_j \rangle=\langle a_j+a^{\dagger}_j \rangle/\sqrt{2}$ for $\alpha=0.01$ and an initial coherent state for the oscillators characterized by $\langle x_1 \rangle=\langle x_2 \rangle=0.5$. Interestingly, we remark that the instantaneous damping rate of the oscillations is temporarily reduced when the oscillators are in phase, as illustrated by the grey areas in Fig.\,\ref{figure1}(a) during which the amplitude of the blue oscillations does not change much, signaling the onset of phase-locking. 

When increasing further $\alpha$, we observe that the pseudo-periods of the two oscillators become identical, and the overall lifetime of the coherence oscillations is greatly increased. This is illustrated in Fig.\,\ref{figure1}(b) for $\alpha=0.16$, where we observe slow-decaying phase-matched oscillations. One can more precisely pinpoint the occurrence of this synchronization mechanism by computing numerically the discrete Fourier transform of $\langle x_j \rangle$(t). The largest peak in Fourier space allows us to determine the dominant frequency in the dynamics $\tilde{\omega}_j$ for both oscillators, that we report in Fig.\,\ref{figure1}(c) for various values of the coupling strength $\alpha$. As can be seen in Fig.\,\ref{figure1}(c), the two frequencies become identical for $\alpha\gtrsim 1.15~10^{-2}$, and they moreover adjust close to the largest original frequency, in accordance with experimental observations (see Fig. 4 in Ref. \cite{Delehaye2015}).  We can reproduce the same procedure for various values of the frequency mismatch $\delta \omega/\omega_0$, and determine the boundary line between the region (I) characterized by unsynchronized decay and region (II) for which phase-locking occurs [see Fig.\,\ref{figure1}(d)]. Unsurprisingly, the critical coupling strength above which phase-locking occurs is seen to grow with the frequency mismatch. In this regime, one also observes a reduced dissipation rate in the system, as can be seen from the long lifetime observed in Fig.\,\ref{figure1}(b).

\begin{figure}[t!]
\includegraphics[scale=0.4]{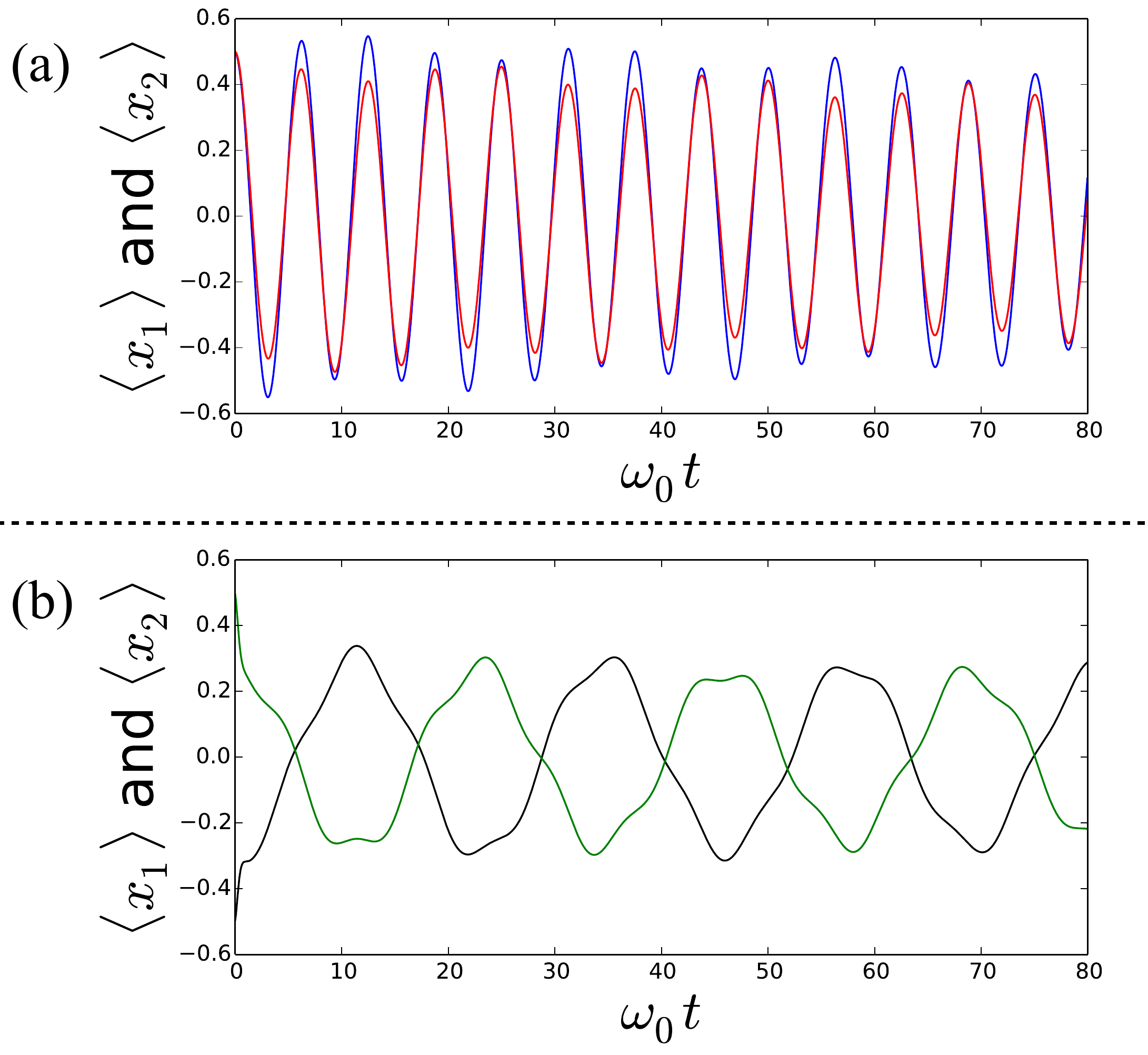}
\caption{Dynamics of $\langle x_1 \rangle$ (blue and green) and $\langle x_2 \rangle$ (red and black) with respect to time for $\delta\omega/\omega_0=0.1$, $\alpha=0.24$, $\omega_c/\omega_0=3$. Initial values are $\langle x_1 \rangle=\langle x_2 \rangle=0.5$ for panel (a) and $\langle x_1 \rangle=-\langle x_2 \rangle=0.5$ for panel (b). }
\label{dissipationless}
\end{figure}

We showed here that the coupling of the two different modes to a common quantum dissipative environment could result in damped oscillations in phase, in analogy to Huyguens' clocks. Next, we will demonstrate the emergence of another dynamical phase with dissipationless (anti)-synchronized oscillations at strong coupling, that do not have any classical analog.


\section{Non-equilibrium quantum phase transition}

Upon further increase of the coupling strength $\alpha$, the dynamics changes with the persistence of seemingly dissipationless oscillations and the emergence of an additional frequency scale. This phenomenon is illustrated in Fig. \ref{dissipationless} (a), which corresponds to $\alpha=0.26$ and $\delta\omega/\omega_0=0.1$ and an initial coherent state for the oscillators characterized by $\langle x_1 \rangle=\langle x_2 \rangle=0.5$. We notably observe that the oscillations of the two oscillators are slowly modulated by an emergent frequency scale $\omega'$. This frequency scale $\omega'$ can be more easily identified in Fig. \ref{dissipationless} (b), for which the initial coherent states are characterized by $\langle x_1 \rangle=-\langle x_2 \rangle=0.5$. We observe in that case a regime with global phase-shifted oscillations with characteristic frequency $\omega'$,  together with small additional higher frequency components.

\begin{figure}[t!]
\includegraphics[scale=0.35]{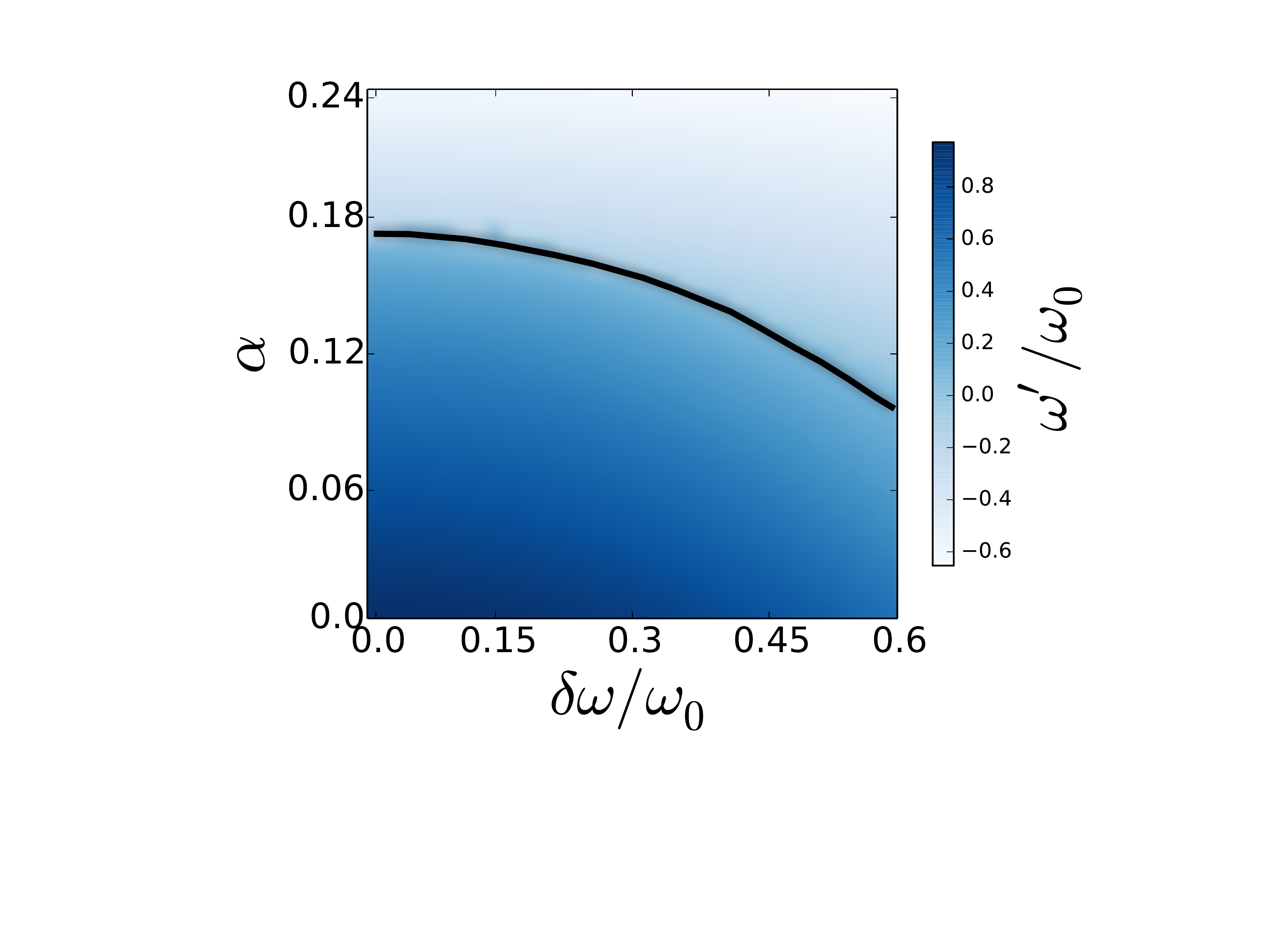}
\caption{Nonequilibrium phase diagram for the dynamics of the oscillators. The color scale shows the value of $\omega'/\omega_0$ obtained from Eq. (\ref{diag}), whose value determine the long time dynamics. The full black line separates the parameter space into two distinct regions, that correspond to trivial steady state $\langle x_j \rangle=0$ (lower part) and long-lived (anti-)synchronized oscillations (upper part). }
\label{phase_diag}
\end{figure}

The appearance of this dissipationless regime results in fact from a nonequilibrium quantum phase transition triggered by the environment\,\cite{Zhang2012,Lin16}. It can be understood by taking the Laplace transform of Eqs. (\ref{eq1}) and (\ref{eq2}), which gives for $U(s)=\int_0^{\infty} dt u(t) e^{-st}$ the following equation
\begin{align}
U(s)=\frac{1}{s \left(1+\frac{\delta \omega^2}{s^2+\omega_0^2}\right)+i\omega_0\left(1-\frac{\delta \omega^2}{s^2+\omega_0^2} \right) +\eta(s)}\label{Laplace},
\end{align}
where $\eta(s)$ is the Laplace transform of $\eta(t)$. The emergence of the long-lived oscillations of Fig.\,\ref{dissipationless} manifests itself by the existence of a pole $\omega'$ of $U$ in Eq.\,(\ref{Laplace}) at negative frequencies, for which there is no possibility of damping through the environment. This condition amounts to the implicit equation  
\begin{align}
\omega'=\omega_0 \left(\frac{\omega'^2+\omega_0^2-\delta\omega^2}{\omega'^2+\omega_0^2+\delta\omega^2}\right)+\Sigma(\omega')\left(\frac{\omega'^2+\omega_0^2}{\omega'^2+\omega_0^2+\delta\omega^2}\right)<0\label{diag},
\end{align}
where $\Sigma(\omega')=2\alpha [\omega' \exp(-\omega'/\omega_c)Ei(\omega'/\omega_c)-\omega_c]$, with $Ei(x)=-\int_{-x}^{\infty} e^{-t}/t$ the exponential integral. The appearance of this dissipativeless pole of $U$ at zero temperature and when the system-environment coupling is above a critical coupling $\alpha_c=(\omega_0^2-\delta \omega^2)/(2\omega_c\omega_0)$ demonstrates a quantum phase transition from dissipation dynamics with damping to localized dynamics. The long-time dynamics of the two highly correlated oscillators in this strong coupling regime is then determined by the frequency scales $\omega_0$, $\delta \omega$, $\omega'$, and the initial state of the system. Interestingly, as illustrated in Fig. \ref{dissipationless}, it can result in either a phase-matched or an out-of-phase dynamics, that furthermore displays genuine quantum correlations, with positive values of the logarithmic negativity\,\cite{Lin16,Vidal02}.
We show in Fig.\,\ref{phase_diag} the nonequilibrium phase diagram of the system with respect to the coupling strength $\alpha$ and the frequency mitsmatch $\delta \omega/\omega_0$. The upper region of parameter space, for which $\omega'/\omega_0<0$, exhibits dissipationless dynamics. Interestingly, one observes that the occurence of the nonequilibrium phase transition occurs for smaller coupling strength $\alpha$ if the two oscillators are originally far detuned. The lower region corresponds to damped dynamics (with oscillations that can be phase-matched or not, see Fig. \ref{figure1} (d)), characterized by steady state values $\langle x_j \rangle=0$.

A similar non-equilibrium transition also occurs for different spectral densities\,\cite{Zhang2012,Lin16}. Let us consider for example the more general spectral function $J(\omega)=\pi \alpha (\omega/\omega_c)^s \omega_c^{1-s} e^{-\omega/\omega_c}$, with $s>0$\,\cite{weiss,Karyn08}. In that case, one finds that the isolated pole at negative frequencies emerges at the critical coupling strength $\alpha_c=(\omega_0^2-\delta \omega^2)/(2\omega_c\omega_0\Gamma(s))$, where $\Gamma$ is the Gamma function.

\section{Conclusion and perspectives}

In this article, we have proposed a simple 
model to explain the spontaneous phase-matching observed in the oscillatory behavior of two coupled cold atomic clouds\,\cite{Delehaye2015}. We have found that this effect can be triggered by the coupling of the relative motion of the two clouds to a quantum dissipative environment, extending results obtained in Refs.\,\cite{these_loic,Karyn_CR2} in the case of two spins 1/2. In that regime, the overall amplitude of the oscillations decay in time to reach the steady-state $\langle x_j \rangle=0$. Upon further increase of the coupling to the environment, we characterized another dynamical transition\,\cite{Zhang2012,Lin16} towards an (anti)-synchronized dissipationless regime. It is important to note that, in all generality, the results presented here could be extended to an arbitrary system-environment coupling. Our results illustrate the richness of dynamical behaviors in quantum systems coupled to non-Markovian structured environments, and show a simple example of spontaneous (anti-)synchronization without external drive in quantum systems. Finally, it would be interesting to explore the possible use of this mechanism to synchronize different clocks, as the spinboson model is a paradigmatic model of dissipation in multiple platforms\,\cite{Karyn:CR}. One could then imagine to operate a clock made of multiple distinct oscillators strongly coupled to the same dissipative environment, to benefit from the coherent frequency locking in their dynamics described in the present manuscript.

We are grateful to Karyn Le Hur and Marion Delehaye for fruitful discussions. We acknowledge support from the Ministry of Economy and Competitiveness
of Spain through the “Severo Ochoa” program for Centres of Excellence in R\&D (SEV-2015-
0522), Fundaci\'{o} Privada Cellex, Fundaci\'{o} Privada Mir-Puig, and Generalitat de Catalunya
through the CERCA program

\bibliographystyle{apsrev4-1}
\bibliography{references}

\end{document}